# Phase-field modeling of solute precipitation and dissolution

Zhijie Xu[1,a)] and Paul Meakin[1,2,3],

1. Idaho National Laboratory Center for Advanced Modeling and Simulation.

2. Physics of Geological Processes, University of Oslo.

3. Multiphase Flow Assurance Innovation Center, Institute for Energy Technology, Kjeller

A phase-field approach to the dynamics of liquid-solid interfaces that evolve due to precipitation and/or dissolution is presented. For the purpose of illustration and comparison with other methods, phase field simulations were carried out assuming first order reaction (dissolution/precipitation) kinetics. In contrast to solidification processes controlled by a temperature field that is continuous across the solid/liquid interface (with a discontinuous temperature gradient) precipitation/dissolution is controlled by a solute concentration field that is discontinuous at the solid/liquid interface. The sharp-interface asymptotic analysis of the phase-field equations for solidification [Karma and Rappel, Phys. Rev. **E57** (1998) 4342] has been modified for precipitation/dissolution processes to demonstrate that the phase-field equations converge to the proper sharp-interface limit. The mathematical model has been validated for a one-dimensional precipitation/dissolution problem by comparison with the analytical solution.

[a)] Electronic mail: zhijie.xu@inl.gov

## I. INTRODUCTION

Precipitation and dissolution at solid-liquid interfaces is of broad scientific interest and practical importance. Significant practical applications include corrosion, etching, the formation of mineral deposits in boilers and heat exchangers, the formation of mineral deposits in oil and water pipes and the formation of gas hydrates in oil and gas pipelines. The dissolution and precipitation of minerals (pressure solution creep[1]) plays an important role in the rheology of the Earth's crust, and the dissolution of minerals provides nutrients for the biosphere. In addition, dissolution and precipitation are important in the formation of economically important mineral deposits and in the formation of a wide range of geological patterns (speleothems, travertine terraces, dissolution scallops, botryoidal precipitation etc.).[2] In practice, mineral precipitation and dissolution often involve complex chemical processes at the solid-liquid interface. To the extent that dissolution and precipitation can be understood in terms of continuously moving boundary problems, phase-field methods provide a robust approach to interface tracking.

Conventional approaches, dating back to Young, Laplace and Gauss in the nineteenth century, are based on the idea that in multiphase systems material properties change discontinuously at interfaces of zero thickness (sharp-interface models). Computational implementation of sharp interface models requires explicit interface tracking, which becomes difficult, error prone or impractical for high dimensionality problems with complicated dynamic geometries, particularly when topological changes such as fragmentation and coallescence occur. In contrast, the phase-field approach, originally developed by van der Waals[3] in the 1800's and by Cahn and Hilliard[4] in the 1950's, is based on the concept of a diffuse interface that can be defined in terms of a density, structure or composition field (the phase field) . The phase field changes smoothly from one phase to the other over an interface zone with a non-zero width, $w$. In numerical applications, the parameters in the phase field equations are selected to ensure that the width of the interface corresponds to several grid cells to achieve a reasonable compromise between accuracy and efficiency. In this manner, numerical difficulties encountered in the sharp-interface model are avoided, and no explicit interface-tracking is needed. The phase field is transported locally with the velocity of the interface and deformation of the phase field is restored by diffusive relaxation (Cahn Hilliard[4]) of conservative phase fields or direct relaxation (Allen-Cahn[5]) of non-conservative fields. Direct simulation of the Cahn-Hilliard equation, including the effects of thermal fluctuations (the Cahn-Hilliard-Cook equation[6]), was initially used to simulate spinodal decomposition[7] and nucleation and growth. Beginning with applications to the solidifications of pure melts,[8-10] the phase field approach has been used to simulate a variety of interface dynamics phenomena (moving boundary problems) including solidification coupled with melt convection,[11,12] two-phase Navier-Stokes flow,[13] grain growth, solid state phase transformations, crack propagation and dislocation dynamics.[14] In most applications, the phase field equations are used to circumvent the difficulty of tracking sharp interfaces, and the phase field equations and/or the parameters used to define the free energy functional do not accurately describe the physics of the interfaces or their dynamics. Here a model for the dynamics of interfaces controlled by precipitation and/or dissolution with phase field interface capturing and the validation of a numerical implementation of the model are described.

## II. SHARP-INTERFACE EQUATIONS AND STEADY STATE SOLUTION

### A. Sharp-interface governing equations

The dynamics of the solid-liquid interface during dissolution or precipitation is a result of the transport of dissolved solid from or to the interface (diffusion in the solid phase is usually small enough to neglect), and the simplest model for solute precipitation/dissolution includes diffusion in the liquid and first order reaction at the liquid-solid interface, without fluid flow. The equation for diffusion of solute in the liquid phase is given by

$$\partial C / \partial t = D\nabla^2 C(\mathbf{x},t)\,, \qquad (1)$$

where $C(\mathbf{x},t)$ is the solute concentration at position $\boldsymbol{x}$ and time $t$, and $D$ is the diffusion coefficient. The equation describing the balance between the solute flux density at the interface, $\Gamma$, and the rate of precipitation or dissolution is

$$D\nabla C|^{+}\bullet\mathbf{n} = k\left(C|^{+} - C_e\right) \text{ on } \Gamma\,, \qquad (2)$$

where $\mathbf{n}$ is the unit vector perpendicular to the interface, $\Gamma$, pointing into the liquid, $C|^{+}$ is the solute concentration in the liquid at the interface (in subsequent equations $|^{+}$ indicates the magnitude of a variable at the liquid side of the interface and $|^{-}$ indicate the magnitude at the solid side of the interface). In Eq. (2), $C_e$ is the solute concentration at equilibrium with the solid, $\nabla C$ is the gradient of the solute concentration, $\mathbf{v}_s$ is the velocity of the interface in the direction normal to the interface and $k$ is the reaction rate coefficient. The magnitude of $\mathbf{v}_s$ is given by

$$v_s = k_c k\left(C|^{+} - C_e\right)\Big/\rho_s\,, \qquad (3)$$

where $\rho_s$ is the density of the solid and $k_c$ is a stoichiometric coefficient of order unity.[15] In Eqs. (2) and (3), the precipitation rate is assumed to be proportional to $C|^+ - C_e$ ($C|^+ > C_e$ implies precipitation and $C|^+ < C_e$ implies dissolution). Simple first order reaction kinetics is rarely, if ever, found in processes such as mineral dissolution and precipitation. However, it has been widely used in simulations to investigate the generic aspect of dissolution and precipitation. Eq. (3) is the local mass conservation condition at an interface that is moving due to solute precipitation or dissolution. The chemical potential at a liquid/solid interface increases as the curvature increases (as the interface become more convex with respect to the liquid) and, consequently, the equilibrium concentration also increases with increasing curvature. This effect (the Gibbs-Thomson effect) is not included in the model. In practice, the Gibbs-Thompson effect is quite small unless the curvature is very large (the radius of curvature is very small). After normalization by the equilibrium concentration, Eqs. (1)-(3) can be rewritten in terms of the dimensionless concentration, $c = \left(C - C_e\right)/C_e$, as

$$\partial c / \partial t = D\nabla^2 c, \tag{4}$$

$$D\nabla c\big|^+ \bullet \mathbf{n} = k\, c\big|^+ \text{ on } \Gamma, \tag{5}$$

and

$$v_s = bk_c k\, c\big|^+ \text{ on } \Gamma, \tag{6}$$

where $c\big|^\pm$ and $\nabla c\big|^\pm$ are the dimensionless normalized solute concentrations and concentration gradients at the interface, and the dimensionless variable $b$ is defined as $b = C_e/\rho_s$.

**B. Sharp-interface steady state solution**

In almost all cases, the width of the liquid solid interface is smooth on macroscopic scales, and the sharp interface model works very well. However, solid-liquid interfaces are often microscopically rough, and the microscopic roughness facilitates precipitation at very small supersaturations, consistent with Eq (3). In these cases a diffuse interface model is used to facilitate interface tracking – not to provide a more accurate physical model for the interface. Under these circumstances, a valid phase field model must give the same results as the sharp interface model in the $w \to 0$ limit where $w$ is the width of the diffuse interface in the phase field model. Consequently, it is useful to first consider the steady state motion of a sharp planar solid-liquid interface since this process can be formulated as a one-dimensional problem, solved analytically, and used to evaluate the phase-field model. In a coordinate system moving with the solid/liquid interface, the constant, time-invariant, concentration profile is given by

$$D\frac{\partial^2 c}{\partial x^2} + v_s \frac{\partial c}{\partial x} = 0, \qquad (7)$$

together with the boundary conditions at the interface (Eqs. (5) and (6)), where $v_s$ is the velocity normal to the interface, in the $+x$ direction, pointing into the liquid. The concentration profile also depends on the far-field boundary condition $c|_{x=+\infty} = c_\infty = (C_\infty - C_e)/C_e$, where $C_\infty$ stands for the far-field solute concentration in the liquid, which corresponds to the far-field temperature, $T_\infty$, in the solidification problem (initially, the solute concentration is $C_\infty$ everywhere in the solution, which extends to infinity). The analytical solution to this steady state propagating problem is

$$v_s = bk_c k\left(c_\infty - \frac{1}{bk_c}\right), \qquad (8)$$

and

$$\begin{cases} c = c_\infty - \dfrac{1}{bk_c}\exp\left(-\dfrac{v_s}{D}x\right) & x > 0 \\ c = 0 & x < 0 \end{cases}. \tag{9}$$

Eq. (9) indicates that the concentration profile in the liquid phase ($x$>0) has an exponential form. The direction of propagation depends on the value of $v_s$ , (either $v_s > 0$ for precipitation from a supersaturated liquid or $v_s < 0$ for dissolution in an undersaturated liquid), and $v_s$ depends on the far-field concentration $C_\infty$ .

## III. PHASE-FIELD MODEL AND ASYMPTOTIC ANALYSIS

### A. Phase-field governing equations for solidifications

Phase-field methods are based on the idea that the free energy associated with the phase field is given by

$$F = \int_V f(\phi) + g(\phi, \nabla\phi) dV , \tag{10}$$

where $\phi$ is the phase field variable, which acts as an indicator function (the value of $\phi$ at any point indicates which phase the point is in), $f(\phi)$ is the free energy density (free energy per unit volume) for a homogeneous system (a system with a uniform phase field) and $g(\phi, \nabla\phi)$ is the contribution to the free energy density due to inhomogeneity (gradients in the phase field), $V$ is the domain occupied by the system of interest and $dV$ is a volume element in the domain. In most applications, particularly those in which the phase field method is used to locate interfaces, the gradient contribution to the free energy functional, $g(\phi, \nabla\phi)$ , is assumed to have the simple form

$$g(\phi,\nabla\phi) = (1/2)\varepsilon^2 \left|\nabla\phi\right|^2 . \tag{11}$$

The basic phase-field equations for solidification of pure melts were first introduced by Langer,[10] and these equations can be derived directly from the free energy functional in the variational form (the VF formulation). The evolution equations for the phase-field variables are described by either the Cahn-Hilliard[4] nonlinear diffusion equation or the Allen-Cahn[5, 16] kinetic equation (relaxation equation), depending on whether the integral of the phase-field variable is conserved or non-conserved. This reduces the interface tracking problem to finding the solution to a set of coupled nonlinear differential equations. In the simplest situation of solidification with an isotropic surface energy, the equations governing the evolution of the phase-field and temperature take the form

$$\tau\frac{\partial\phi}{\partial t} = -\frac{\delta F}{\delta\phi} = \left\{\varepsilon^2\nabla^2\phi - \frac{\partial f\left(\phi,T\right)}{\partial\phi}\right\}, \tag{12}$$

and

$$\frac{\partial T}{\partial t} = D_T\nabla^2 T + A\frac{\partial\phi}{\partial t}, \tag{13}$$

where $\tau$ is a positive characteristic time reflecting the atom mobility. When the phase-field is used as an indicator function to locate the interface, the parameter $\tau$ in Eq. (12) is selected for computational convenience and accuracy. The coefficient $\varepsilon$ in Eqs. (11) and (12) has the dimension of length, and it is closely related the interface thickness. $T$ is the temperature field, $D_T$ the thermal diffusivity and $A$ is a constant obtained from the interface boundary condition. We adopt the convention $\phi = -1$ and $\phi = 1$ to indicate the solid and liquid phases in our phase-field model. Points with $-(1-\delta) < \phi < (1-\delta)$ lie in the diffuse interface zone. In both the theoretical work and the numerical applications, $\phi$ is set to +1 or -1 outside of the interface zone. In phase-field models the values of the physical quantities vary continuously, and there is no clear demarcation between the bulk phase and the interface zone. From a practical point of view, $\delta$ is small, but not too small (a value on the order of $10^{-2}$ would be reasonable in numerical work). To apply the phase-field approach to solidification, the heat diffusion equation, Eq. (13), was modified by adding a heat source term to account for the heat produced (latent heat of solidification) as the interface moves, and the free energy functional takes the general form

$$F = \int_V \left\{ \frac{1}{2}\varepsilon^2 \left|\nabla\phi\right|^2 + f\left(\phi, T\right) \right\} dV \,. \tag{14}$$

The gradient term (the first term on the right hand side of Eq. (14)) is related to the excess interfacial energy density (the energy needed to create extra surface area). The function $f\left(\phi, T\right)$ is assumed to have a double-well potential form with two minima corresponding to the solid and liquid phases. In addition, $f\left(\phi, T\right)$ couples the temperature field $T$ with the phase field $\phi$ to account for the effect of temperature on the free energy density.

## B. Phase-field equations for solute precipitation and dissolution

The formulation of basic phase field equations for dissolution/precipitation can be based on the similarities and differences between solidification (described by Eqs. (12) and (13)) and precipitation/dissolution. Phase field interface tracking or capturing is based on the idea that the phase field profile evolves towards a constant form across the moving interface under advection (transport with the interface) and control of the Cahn-Hilliard equation or Allen-Cahn equation. However, the relatively rapid convergence of the phase field in the interface zone is accompanied by curvature driven motion of the interface. If a physical free energy functional is used to simulate interface dynamics, the curvature driven interface motion is an important part of the process, which is related to the excess interfacial free energy. On the other hand, if the phase field is used for interface tracking or capturing, the curvature driven interface motion is a significant source of error, which can be counteracted by replacing Eq. (12) with [17-19]

$$\tau \frac{\partial \phi}{\partial t} = \varepsilon^2 \nabla^2 \phi - \frac{\partial f}{\partial \phi} - \varepsilon^2 \left| \nabla \phi \right| \kappa . \tag{15}$$

In Eq. (15) $\kappa$ is the curvature, which can be calculated from $\kappa = \nabla \cdot (\nabla \phi / |\nabla \phi|) = \nabla \cdot \bar{\mathbf{n}}$, where $\bar{\mathbf{n}}$ is the unit vector in the direction of the phase filed gradient. In the sharp-interface model, the curvature $\kappa$ is defined only on the interface, and $\kappa = \nabla \cdot \bar{\mathbf{n}}$, where $\bar{\mathbf{n}}$ is the unit vector perpendicular to the interface, pointing into the liquid phase. The last term on the right-hand-side of Eq. (15) counteracts the curvature driven interface motion to leading order in the small parameter $\varepsilon\kappa$.

To determine the sharp interface limit, Eq. (15) is transformed to a set of local orthogonal curvilinear coordinates ($s_1, s_2$ and $n$) that move with the interface ($s_1$ and $s_2$ are the coordinates along the two principal directions along the interface, and $n$ is the coordinate in the direction perpendicular to the interface).[20] Transformation to the moving curvilinear coordinate system ($\nabla^2 = \partial^2 / \partial n^2 + (\partial / \partial n)\kappa,$ ignoring higher order terms in $\varepsilon\kappa$) introduces a new curvature related term that eliminates the curvature correction in Eq. (15), and Eq. (15) can be expressed as

$$\tau \frac{\partial \phi}{\partial t} = \varepsilon^2 \frac{\partial^2 \phi}{\partial n^2} - \frac{\partial f}{\partial \phi} . \tag{16}$$

This equation leads to a stationary hyperbolic tangent interface profile across the interface.[19]

In Eq. (13), the term $A\partial\phi/\partial t$ describes the liberation of latent due to the solidification process. In precipitation or dissolution, dissolved solid (solute) is removed from or released to the liquid as the solid/liquid interface advances or recedes, and the corresponding equation for the concentration field is:

$$\frac{\partial c}{\partial t} = D\nabla^2 c + A_1 \frac{\partial \phi}{\partial t}. \tag{17}$$

The second term on the right-hand-side of Eq. (17) provides a solute sink or source (depending on the direction in which the interface is moving). In the phase field model, this is a diffuse source or sink localized in the vicinity of the interface ($\partial\phi / \partial t$ is negligibly small outside of the diffuse interface region). Since $\phi$ varies from -1 to 1, it follows that $A_1 = \rho_s / 2C_e k_c$ in the $\varepsilon \to 0$ limit.

In the dissolution/precipitation problem, additional terms are needed to ensure that the boundary conditions given in Eqs. (5) and (6) are satisfied in the sharp-interface limit. One of these terms must be a sink or source term to account for the fact that during dissolution (or precipitation) solid with a density of $\rho_s$ is converted into solute with a non-zero concentration (or solute with a non-zero concentration is converted into solid) and the other term must generate the solute concentration discontinuity across the interface. These considerations, and the requirement that the solution of the phase-field equations must converge to the solution of the corresponding sharp-interface problem as $\varepsilon \to 0$ lead to the combination of Eq. (15) with

$$\frac{\partial c}{\partial t} = D\nabla^2 c + A_1 \frac{\partial \phi}{\partial t} + A_2 \frac{\partial\phi/\partial t}{|\nabla\phi|}\left( D\nabla^2\phi - \frac{\partial \phi}{\partial t} \right). \tag{18}$$

The second term on the right-hand-side of Eq. (18) acts as a net source or sink of solute corresponding to the discontinuity in the solute concentration gradient across the interface, while the third term acts as net sources or sinks of solute coming from the discontinuity in the solute concentration across the interface. There is no term analogous to the third term in Eq. (18) for solidification problem as the temperature field is continuous across the interface.[20] The phase field precipitation/dissolution model must converge to the correct sharp interface limit as $\varepsilon \to 0$.

In the precipitation/dissolution model, the free energy density is assumed to have the form

$$f(\phi, c) = f_1(\phi) + \lambda f_2(\phi) c \, , \tag{19}$$

and this couples the phase field with the concentration field. A double-well function with minima at $\phi = -1$ and $\phi = 1$ is used for $f_1(\phi)$. In Eq. (19), $f_2(\phi)$ is required to be an odd function of $\phi$ with zero gradients at $\phi = -1$ and $\phi = 1$. This form for the homogeneous free energy density has the advantage that $f(\phi, c)$ has two minima at $\phi = \pm 1$, independent of the value of *c*. To facilitate the analysis needed to establish the sharp-interface limit, the simple functional forms $f_1(\phi) = -\phi^2/2 + \phi^4/4$ and $f_2(\phi) = \phi - \phi^3/3$ are used. The dimensionless parameter $\lambda$ is used to control the strength of coupling between the concentration field, *c*, and the phase-field, $\phi$. The two constants, $A_1$ and $A_2$ in Eq. (18) are determined by the sharp-interface boundary conditions (Eqs. (5) and (6)) and the requirement that the phase field model must convergence to the correct sharp interface limit. In the ($s_1, s_2 \;\; n$) curvilinear coordinate system moving with the interface at an instantaneous velocity of $v_n$, with the normal direction pointing into the liquid, Eq. (18) can be rewritten as,

$$\frac{D}{v_n}\frac{\partial^2 c}{\partial n^2}+\left(\frac{D\kappa}{v_n}+1\right)\frac{\partial c}{\partial n}=\left(A_1+A_2 v_n\right)\frac{\partial \phi}{\partial n}+A_2 D\left(\frac{\partial^2 \phi}{\partial n^2}+\kappa\frac{\partial \phi}{\partial n}\right), \tag{20}$$

where $n$ is the coordinate along the interface normal. Integration of both sides of Eq. (20) across the interface in the sharp-interface limit in which the phase field $\phi$ becomes a step function and velocity normal to the interface becomes the sharp-interface velocity, $v_s$, gives

$$\frac{D}{v_s}\frac{\partial c}{\partial n}\bigg|_{0^-}^{0^+}+\left(\frac{D\kappa}{v_s}+1\right)c\Big|_{0^-}^{0^+}=\left(A_1+A_2 v_s\right)\phi\Big|_{0^-}^{0^+}+A_2 D\left(\frac{\partial \phi}{\partial n}\bigg|_{0^-}^{0^+}+\kappa\phi\Big|_{0^-}^{0^+}\right). \tag{21}$$

In the sharp interface limit, the two terms on the left-hand-side can be evaluated from Eqs. (5) and (6) as:

$$\frac{D}{v_s}\frac{\partial c}{\partial n}\bigg|_{0^-}^{0^+}=\frac{1}{bk_c} \text{ and } c\Big|_{0^-}^{0^+}=\frac{v_s}{bk_c k}. \tag{22}$$

The constants $A_1$ and $A_2$ can be determined from the boundary conditions for the phase field, $\phi$, ($\phi|_{n=0^\pm}=\pm 1$, $\partial\phi/\partial n|_{n=0^\pm}=0$) and the dimensionless concentration field, $c$, in the limit in which $\phi$ is the step function from -1 to 1 and $c$ is the exact solution of the sharp-interface model. This leads directly to the results $A_1=1/(2bk_c)$, or $A_1=\rho_s/2C_e k_c$ and $A_2=1/(2bk_c k)$. Integrating Eq. (20) twice across the interface as $\phi$ approaches the step function and $c$ approaches the exact solution to the sharp-interface model gives

$$\frac{D}{v_s}c\Big|_{0^-}^{0^+}+\left(\frac{D\kappa}{v_s}+1\right)\int_{0^-}^{0^+} c\cdot dn=\left(A_1+A_2 v_s+A_2 D\kappa\right)\int_{0^-}^{0^+}\phi\cdot dn+A_2 D\phi\Big|_{0^-}^{0^+}. \tag{23}$$

In the sharp interface limit, a step function can be used for the order parameter profile, the integrals of $\phi$ and $c$ across the interface in Eq. (23) are zero, and it can easily be shown that $A_2=1/(2bk_c k)$ using the sharp-interface equation for the interface velocity (Eq. (6)), which implies that $c\big|_{0^-}^{0^+} = v_s/(bk_c k)$. The phase-field equation for precipitation/dissolution can then be written as:

$$\tau\frac{\partial\phi}{\partial t} = \left\{\varepsilon^2\nabla^2\phi - \frac{\partial f_1}{\partial\phi} - \lambda c\frac{\partial f_2}{\partial\phi}\right\} - \varepsilon^2\left|\nabla\phi\right|\kappa, \tag{24}$$

and substitution of the expressions obtained for $A_1$ and $A_2$ into Eq. (18) for the evolution of the concentration field gives

$$\frac{\partial c}{\partial t} = D\nabla^2 c + \frac{1}{2bk_c}\frac{\partial\phi}{\partial t}\left(1 + \frac{D\nabla^2\phi - \partial\phi/\partial t}{k\left|\nabla\phi\right|}\right). \tag{25}$$

Eqs. (24) and (25) satisfy the boundary conditions in the sharp-interface limit. These equations provide only a phenomenological description of the underlying precipitation/dissolution processes at the solid-liquid interface. However, valid phase-field equations must quantitatively reduce to the corresponding free-boundary problem in the sharp-interface limit. This has been demonstrated using asymptotic analysis for the solidification problem by Langer,[10] Caginalp[21] and Karma and Rappel.[20] The sharp-interface model neglects microscopic details at the interface; however its parameters (for example, the reaction rate constant *k*) can be quantitatively related to the phase-field microscopic parameters by the formal asymptotic analysis.

### C. Asymptotic analysis of the phase-field equations and the thin-interface limit

To facilitate the asymptotic analysis, the phase-field Eqs. (24) and (25) were first rewritten in a dimensionless form. To accomplish this, a characteristic length scale $l_c$ (the solute diffusion length) and a characteristic time scale $t_c = l_c^2/D$ are used as the units of length and time.[20] With these length and time units, the characteristic interface velocity is $v_c = D/l_c$ , and the dimensionless phase-field equations are

$$\eta\gamma^2 \frac{\partial \phi}{\partial t} = \gamma^2 \nabla^2 \phi - \frac{\partial f_1}{\partial \phi} - \lambda c \frac{\partial f_2}{\partial \phi} - \gamma^2 \left|\nabla \phi\right| \kappa , \tag{26}$$

and

$$\frac{\partial c}{\partial t} = \nabla^2 c + \frac{1}{2bk_c}\left(1 + \frac{\nabla^2 \phi - \partial \phi/\partial t}{k\left|\nabla \phi\right|}\right)\frac{\partial \phi}{\partial t} , \tag{27}$$

where $\gamma = \varepsilon/l_c = \varepsilon v_c/D$ and $\eta = D\tau/\varepsilon^2$ . (28)

Here, $\gamma$ is a small parameter (the interface thickness, ε, is small compared with the diffusion length, $l_c$ , in the phase field model and ε, becomes negligible compared with $l_c$ in the sharp interface limit). The dimensionless parameter, $\gamma$ , can also be interpreted as the 'interface Peclet number'. Asymptotic analysis was then performed by seeking the solution in the inner region (the diffuse interface region where $\phi$ varies rapidly in the direction perpendicular to the interface) and the solution in the outer region (the bulk phase where $\phi$ varies slowly), and matching them to each other. The outer region solutions (outer solutions) can be easily identified as $\phi^o = \pm 1$ in the liquid and solid phases. The inner region solutions (inner solutions) can be expressed as a power series in the small parameter $\gamma$ ,

$$\phi^i = \phi_0^i + \gamma\phi_1^i + \gamma^2\phi_2^i + \ldots , \tag{29}$$

and

$$c^i = c_0^i + \gamma c_1^i + \gamma^2 c_2^i + \ldots , \tag{30}$$

where the superscripts $i$ and $o$ the stand for inner and outer solutions. An inner variable, $\xi = n/\gamma$, is defined along the interface normal pointing into the liquid phase and Eqs. (26) and (27) are rewritten in the local orthogonal curvilinear coordinate system ($s_1$, $s_2$, $\xi$),

$$\frac{\partial^2 \phi}{\partial \xi^2} + \eta \gamma v_n \frac{\partial \phi}{\partial \xi} = \frac{\partial f_1}{\partial \phi} + \lambda c \frac{\partial f_2}{\partial \phi}, \tag{31}$$

$$\frac{\partial^2 c}{\partial \xi^2} + \gamma(\kappa + v_n)\frac{\partial c}{\partial \xi} = \gamma v_n (\alpha + \beta + \theta)\frac{\partial \phi}{\partial \xi} + \beta \frac{\partial^2 \phi}{\partial \xi^2}, \tag{32}$$

where $\alpha = 1/(2bk_c)$, $\beta = v_n/(2bk_c k)$ and $\theta = \kappa/(2bk_c k)$. (33)

By substituting the power series (29) and (30) into the phase-field Eqs. (31) and (32), substituting the expressions $f_1(\phi) = -\phi^2/2 + \phi^4/4$ and $f_2(\phi) = \phi - \phi^3/3$, for $f_1$ and $f_2$ into Eq. (31), and retaining only the leading terms (the terms of order zero in $\gamma$), the relationships

$$\frac{\partial^2 \phi_0^i}{\partial \xi^2} + (\phi_0^i - \lambda c_0^i)(1 - \phi_0^{i2}) = 0, \tag{34}$$

and

$$\frac{\partial^2 c_0^i}{\partial \xi^2} - \beta \frac{\partial^2 \phi_0^i}{\partial \xi^2} = 0, \tag{35}$$

can be obtained. The solution of these equations is

$$\phi_0^i = \tanh\left(\xi\sqrt{(1-\lambda\beta)/2}\right), \tag{36}$$

and

$$c_0^i = \beta\phi_0^i = \beta \tanh\left(\xi\sqrt{(1-\lambda\beta)/2}\right). \tag{37}$$

Similarly, the terms that are first order in $\gamma$ give rise to the equations,

$$\frac{\partial^2 \phi_1^i}{\partial \xi^2} + \eta v_n \frac{\partial \phi_0^i}{\partial \xi} = -\phi_1^i(1 - 3\phi_o^{i2}) + \lambda c_1^i (1 - \phi_0^{i2}) - 2\lambda c_0^i \phi_0^i \phi_1^i, \tag{38}$$

and

$$\frac{\partial^2 c_1^i}{\partial \xi^2}+\left(\kappa+v_n\right)\frac{\partial c_0^i}{\partial \xi}=v_n\left(\alpha+\beta+\theta\right)\frac{\partial \phi_0^i}{\partial \xi}+\beta\frac{\partial^2 \phi_1^i}{\partial \xi^2}\,. \tag{39}$$

Substituting the solutions of the leading order equations (Eqs. (36) and (37)) into the first order Eqs. (38) and (39) gives

$$\left\{\frac{\partial^2}{\partial \xi^2}+\left(1-3\phi_o^{i2}+2\lambda\beta\phi_0^{i2}\right)\right\}\phi_1^i=-\eta v_n\frac{\partial \phi_o^i}{\partial \xi}+\lambda c_1^i\left(1-\phi_o^{i2}\right), \tag{40}$$

and

$$\frac{\partial^2 c_1^i}{\partial \xi^2}=\alpha v_n\frac{\partial \phi_0^i}{\partial \xi}+\beta\frac{\partial^2 \phi_1^i}{\partial \xi^2}\,. \tag{41}$$

By direct integration, the solution to Eq. (41) is

$$c_1^i=\beta\phi_1^i+\alpha v_n\int_0^{\xi}\phi_0^i\left(\xi'\right)d\xi'+B_1\xi+B_2\,. \tag{42}$$

The two integration constants $B_1$ and $B_2$ can be determined by matching the inner and outer solutions. By substituting the solution for $c_1^i$ (Eq. (42)) into Eq. (40) and rearranging, the equation

$$L\phi_1^i=\left\{\frac{\partial^2}{\partial \xi^2}+\left(1-\lambda\beta\right)\left(1-3\phi_o^{i2}\right)\right\}\phi_1^i=RHS\,, \tag{43a}$$

with

$$RHS=-\eta v_n\frac{\partial \phi_o^i}{\partial \xi}+\lambda\left(1-\phi_o^{i2}\right)\left(\alpha v_n\int_0^{\xi}\phi_0^i\left(\xi'\right)d\xi'+B_1\xi+B_2\right), \tag{43b}$$

is obtained. It can easily be demonstrated that $\partial\phi_0^i/\partial\xi$ satisfies the homogeneous equation $L\left(\partial\phi_0^i/\partial\xi\right)=0$. Therefore, the right hand side of Eq. (43a) (RHS) and $\partial\phi_0^i/\partial\xi$ must satisfy the condition that $\int_{-\infty}^{+\infty}\left(\partial\phi_0^i/\partial\xi\right)\cdot RHS\cdot d\xi=0$ for a solution of Eq. (40) to exist. The RHS of Eq. (43a) (i.e. Eq. (43b)) can be broken into four parts, and we have:

$$-\eta v_n I_1+\lambda B_2 I_2+\lambda\alpha v_n I_3+\lambda B_1 I_4=0, \tag{44}$$

where

$$I_1=\int_{-\infty}^{+\infty}\left(\partial\phi_0^i/\partial\xi\right)^2 d\xi=2\sqrt{2\left(1-\lambda\beta\right)}\Big/3,$$

$$I_2=\int_{-\infty}^{+\infty}\left(1-\phi_o^{i2}\right)\left(\partial\phi_0^i/\partial\xi\right)d\xi=4/3,$$

$$I_3=\int_{-\infty}^{+\infty}\left(1-\phi_o^{i2}\right)\left(\partial\phi_0^i/\partial\xi\right)\int_0^{\xi}\phi_o^i\left(\xi'\right)d\xi' d\xi=\frac{\left(5/9-\left(2\ln 2\right)/3\right)2\sqrt{2}}{\sqrt{1-\lambda\beta}},$$

$$I_4=\int_{-\infty}^{+\infty}\left(1-\phi_o^{i2}\right)\left(\partial\phi_0^i/\partial\xi\right)\xi\cdot d\xi=0. \tag{45}$$

The integral $I_4$ is zero because $\left(\partial\phi_0^i/\partial\xi\right)$ is a even function of $\xi$. Finally, the expression

$$B_2=\left\{\eta I_1/\left(\lambda I_2\right)-\alpha I_3/I_2\right\}v_n, \tag{46}$$

is obtained for the second integration constant, $B_2$, from Eq. (44). The outer solutions on both sides of the solid-liquid interface can be expanded up to first order in the matching regions,

$$c^o=c^o\Big|^{\pm}+\partial c^o/\partial x\Big|^{\pm}\cdot x, \tag{47}$$

where $c^o\,|^+$ and $c^o\,|^-$ are the outer solutions at the interface on the liquid and solid sides of the diffuse interface. The matching conditions require both the slopes and magnitudes of the inner and outer solutions to be equal as the interface is approached ($\lim_{x\to 0^{\pm}}\left(\partial c^o/\partial x\right)=\lim_{\xi\to\pm\infty}\left(\partial c^i/\partial x\right)$ and $c^o=\lim_{\xi\to\pm\infty}c^i$). The first integration constant $B_1$ in Eq. (42) can be found by substituting $c^o$ and $c^i=c_0^i+\gamma c_1^i$ into the slope matching condition ($\lim_{x\to 0^{\pm}}\left(\partial c^o/\partial x\right)=\lim_{\xi\to\pm\infty}\left(\partial c^i/\partial x\right)$) to obtain the result

$$B_1 = \partial c^o / \partial x\Big|^{+} - \alpha v_n = \partial c^o / \partial x\Big|^{-} + \alpha v_n \, . \tag{48}$$

Substitution of Eq. (48) for $B_1$ and Eq. (46) for $B_2$ into the first order solution for $c_1^i$ (Eq. (42)) leads to the result that

$$\lim_{\xi \to \pm\infty} c_1^i = \alpha v_n \int_0^{\pm\infty} \left( \phi_0^i \left( \xi' \right) \mp 1 \right) d\xi' + B_2 + \frac{\partial c}{\partial x}\bigg|^{\pm} \cdot \xi \, . \tag{49}$$

The inner solution in the matching region can then be written as

$$\lim_{\xi \to \pm\infty} c^i = \pm\beta + \gamma \alpha v_n \int_0^{\pm\infty} \left( \phi_0^i \left( \xi' \right) \mp 1 \right) d\xi' + B_2 + \frac{\partial c}{\partial x}\bigg|^{\pm} \cdot x \, , \tag{50}$$

and by equating Eqs. (50) and (47) (value matching), the outer solution

$$c^o\Big|^{\pm} = \pm\beta + \gamma v_n \left\{ \alpha \left( I_5 - I_3 / I_2 \right) + \eta / \lambda \cdot I_1 / I_2 \right\}, \tag{51}$$

is obtained at both sides of the interface, where $I_5 = \int_0^{\pm\infty} \left( \phi_0^i \mp 1 \right) d\xi = -\sqrt{2} \ln 2 / \sqrt{1 - \lambda\beta}$ . It can be easily verified that the concentration gradient discontinuity at the interface is $2\alpha v_n$ (Eq. (48)) and that the concentration discontinuity in the sharp-interface limit is $2\beta$ (Eq. (51)). The phase-field equations must reduce to the corresponding sharp-interface equations in the small interface thickness limit. By using the fact that $c^o\Big|^{-} = 0$ on the solid phase side of the interface and by substituting the definitions for $\eta$ and $\gamma$ given in Eq. (28) and the values of the integrals $I_1$-$I_4$ in Eq. (45), into Eq. (51), the asymptotic solution

$$v_n \approx v_a \equiv \frac{\beta \sqrt{1 - \lambda\beta}}{\dfrac{\sqrt{2}}{2} \dfrac{\tau}{\lambda\varepsilon} \left( 1 - \lambda\beta \right) - \dfrac{5\sqrt{2}}{6} \dfrac{\alpha\varepsilon}{D}}, \tag{52}$$

is obtained for the normal velocity of the interface. Eq. (52) should give exact agreement with the relationship $v_n \equiv \beta k / \alpha$ obtained from the definitions of $\alpha$ and $\beta$ given by Eq. (33)

as $\beta \to 0$, and in this limit the interface velocity, $v_n$ also approaches zero. In this limit, the relationship

$$k = \frac{1}{\frac{\sqrt{2}}{2\alpha} \cdot \frac{\tau}{\lambda\varepsilon} - \frac{5\sqrt{2}}{6} \frac{\varepsilon}{D}}, \tag{53}$$

between the reaction rate constant *k*, a macroscopically measurable quantity, and the phase-field microscopic parameters is obtained. Eq. (53) provides a basis for relating the phase-field model to the corresponding sharp-interface model. It implies that the convergence of the phase-field model to the corresponding free boundary problem can be achieved by decreasing the phase-field microscopic parameters $\varepsilon$, $\tau$ and $\lambda$ with a fixed reaction rate constant *k* until the phase-field result becomes independent of these microscopic phase-field parameters. Eq. (53) indicates that in order to keep the reaction rate constant, $\tau$ must be related to $\varepsilon$ and $\lambda$ and *k* through the equation

$$\tau = \alpha\lambda\varepsilon\left(\frac{5\varepsilon}{3D} + \frac{\sqrt{2}}{k}\right). \tag{54}$$

## IV. ONE-DIMENSIONAL STEADY STATE COMPARISON OF SHARP-INTERFACE AND PHASE-FIELD MODELS

In order to validate the connections between the phase-field model and the corresponding sharp-interface problem, the coupled phase-field equations must be solved. The steady state phase-field equations are simply:

$$\varepsilon^2 \frac{\partial^2 \phi}{\partial x^2} + \tau v_n \frac{\partial \phi}{\partial x} + (\phi - \lambda c)(1 - \phi^2) = 0, \tag{55}$$

and

$$D\frac{\partial^2 c}{\partial x^2}+v_n\frac{\partial c}{\partial x}=(\alpha+\beta)v_n\frac{\partial \phi}{\partial x}+\beta D\frac{\partial^2 \phi}{\partial x^2}, \quad (56)$$

in a one-dimensional coordinate system moving with a constant velocity $v_n$ (or a planar surface moving in three dimensions with a velocity constant velocity $v_n$ in the direction perpendicular to the interface). Since the analytical solution of the sharp-interface problem has the exponential form $\exp(-v_s x/D)$ (Eq. (9)), the steady state phase-field equations are expressed in terms of the dimensionless length $x_1 = x\,v_n/D$ to give

$$\frac{\partial^2 \phi}{\partial x_1^2}+\frac{\tau D}{\varepsilon^2}\frac{\partial \phi}{\partial x_1}=\left(\frac{D}{\varepsilon v_n}\right)^2(\lambda_1 c_1-\phi)(1-\phi^2), \quad (57)$$

and

$$\frac{\partial^2 c_1}{\partial x_1^2}+\frac{\partial c_1}{\partial x_1}=(\alpha_1+\beta_1)\frac{\partial \phi}{\partial x_1}+\beta_1\frac{\partial^2 \phi}{\partial x_1^2}, \quad (58)$$

where $c_1 = c/c_\infty$ , $\alpha_1 = \alpha/c_\infty$ , $\beta_1 = \beta/c_\infty$ and $\lambda_1 = \lambda c_\infty$ . The original interface tracking problem is then reduced to seeking the solutions of the coupled differential Eqs. (57) and (58) for $\phi$, $c_1$, $v_n$ and $\beta_1$ subject to the normalized far-field boundary conditions ($c_1|_{x=+\infty}=1$ and $c_1|_{x=-\infty}=0$) for any given $\alpha_1$. This boundary value problem can be solved using standard numerical schemes. For comparison, the solutions of corresponding sharp-interface analytical are

$$c_s = 1-2\alpha_1\exp(-x_1) \text{ and } v_s = k\omega, \quad (59)$$

where $\omega=\beta_1/\alpha_1$. Fig.1 show the normal interface velocity as a function of $\omega$ obtained from the phase-field and sharp-interface models. As expected, the phase-field velocity $v_n$ is in good agreement with the asymptotic solution $v_a$ in Eq. (52). In the sharp-interface limit with $\omega$ or $v_n \to 0$ (small parameter $\gamma \to 0$ ), both the asymptotic and phase-field solutions reproduce the sharp-interface limit (Eq. (59)) solution, $v_s = k\omega$ , where $k$ is the kinetic reaction rate. Alternatively, for a given $\omega$, the sharp-interface limit can be recovered with the phase-field microscopic parameters $\varepsilon, \tau, \lambda \to 0$ while the rate constant $k$ is kept constant according to Eq. (54). By decreasing $\varepsilon$ and $\lambda$ simultaneously, a comparison between the sharp-interface steady-state concentration profile $c_s$ and velocity $v_s$ (Eq. (59)) and the phase-field solutions of $c_1$ and $v_n$ are presented in Fig.2 and Fig.3. The phase-field solutions for $\phi$ is also presented in Fig.2. As expected, the interface becomes narrower, represented by the variation of $\phi$, as $\varepsilon$ and $\lambda$ decrease, while the concentration $c_1$ and interface velocity $v_n$ approaches the sharp-interface solution $c_s$ and $v_s$ asymptotically. These comparisons validated the phase-field model for precipitation and dissolution.

## V. CONCLUSIONS

A phase-field approach to modeling solute precipitation and dissolution at solid-liquid interfaces has been developed. The interface thickness was assumed to be finite but small compared with the length scale of the growth pattern. In contrast to the standard phase-field approach for solidification of pure melts, the approach described here introduces additional terms in the modified solute diffusion equation to account for the solute concentration discontinuity at the liquid-solid interface. Using a detailed asymptotic analysis, the connections between the sharp-interface and phase-field models were established by relating the reaction rate constant $k$ to the microscopic phase-field parameters (Eq. (53)). This ensures that the phase-field model will converge to the corresponding free-boundary problem. The model was validated by a one-dimensional investigation of interface motion due to solute precipitation.

## ACKNOWLEDGMENTS

This work was supported by the U.S. Department of Energy, Office of Science Scientific Discovery through Advanced Computing Program. The Idaho National Laboratory is operated for the U.S. Department of Energy by the Battelle Energy Alliance under Contract DE-AC07-05ID14517.

FIG. 1. Dependence of the phase-field interface normal velocity $v_n$ on $\omega = \beta_1 / \alpha_1$ from phase-field Eqs. (57) and (58) for steady state planar interface propagation compared with the sharp-interface limit $v_s$ predicted by Eq. (59) and the asymptotic solution $v_a$ predicted by Eq. (52). The phase-field results converge to the sharp-interface limit as $v_n$ approaches 0.

FIG. 2. Numerically calculated phase-field variable $\phi$ (the interface profile) and solute concentration profile $c$ obtained from phase-field Eqs. (57) and (58) for a planar steady-state propagating interface compared with the sharp-interface limit predicted by Eq. (59) for three values of the microscopic phase-field parameters $\varepsilon$ and $\lambda$, with $\varepsilon = \lambda$. The phase-field results converge to the corresponding sharp-interface solution as $\varepsilon$ and $\lambda$ approach 0 while the rate constant $k$ remains constant.

FIG. 3. The phase-field velocity $v_n$ calculated numerically from phase-field Eqs. (57) and (58) for steady-state propagation of a sharp-interface compared to the sharp-interface limit predicted by Eq. (59) for various values of the microscopic phase-field parameters $\varepsilon$ and $\lambda$. All velocities were normalized by the sharp-interface limit velocity, $v_s$. The phase-field results converge to the corresponding sharp-interface limit as $\varepsilon$ and $\lambda$ (numerically equal) approach 0 while the rate constant, $k$, is constant.

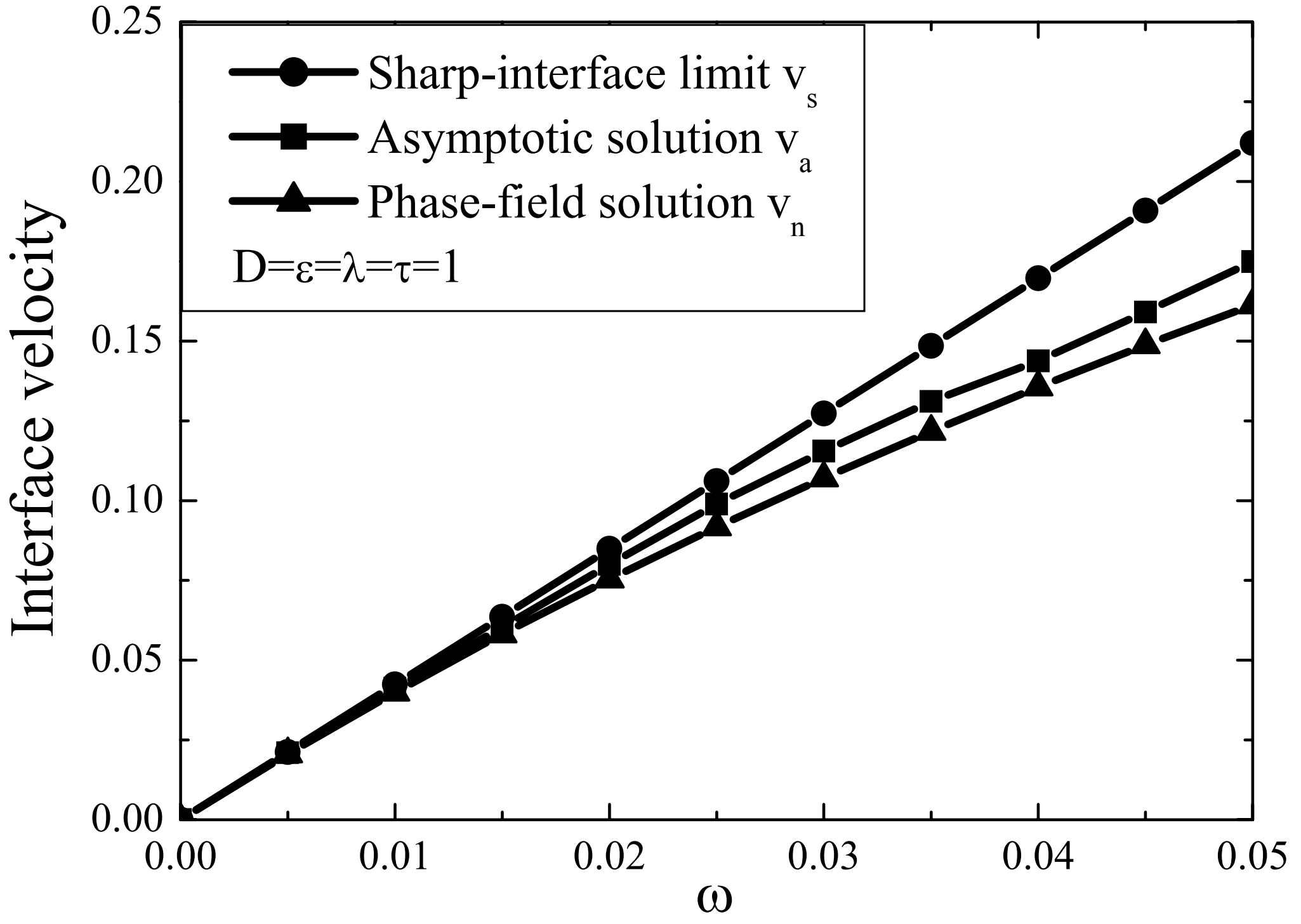

FIG. 1.

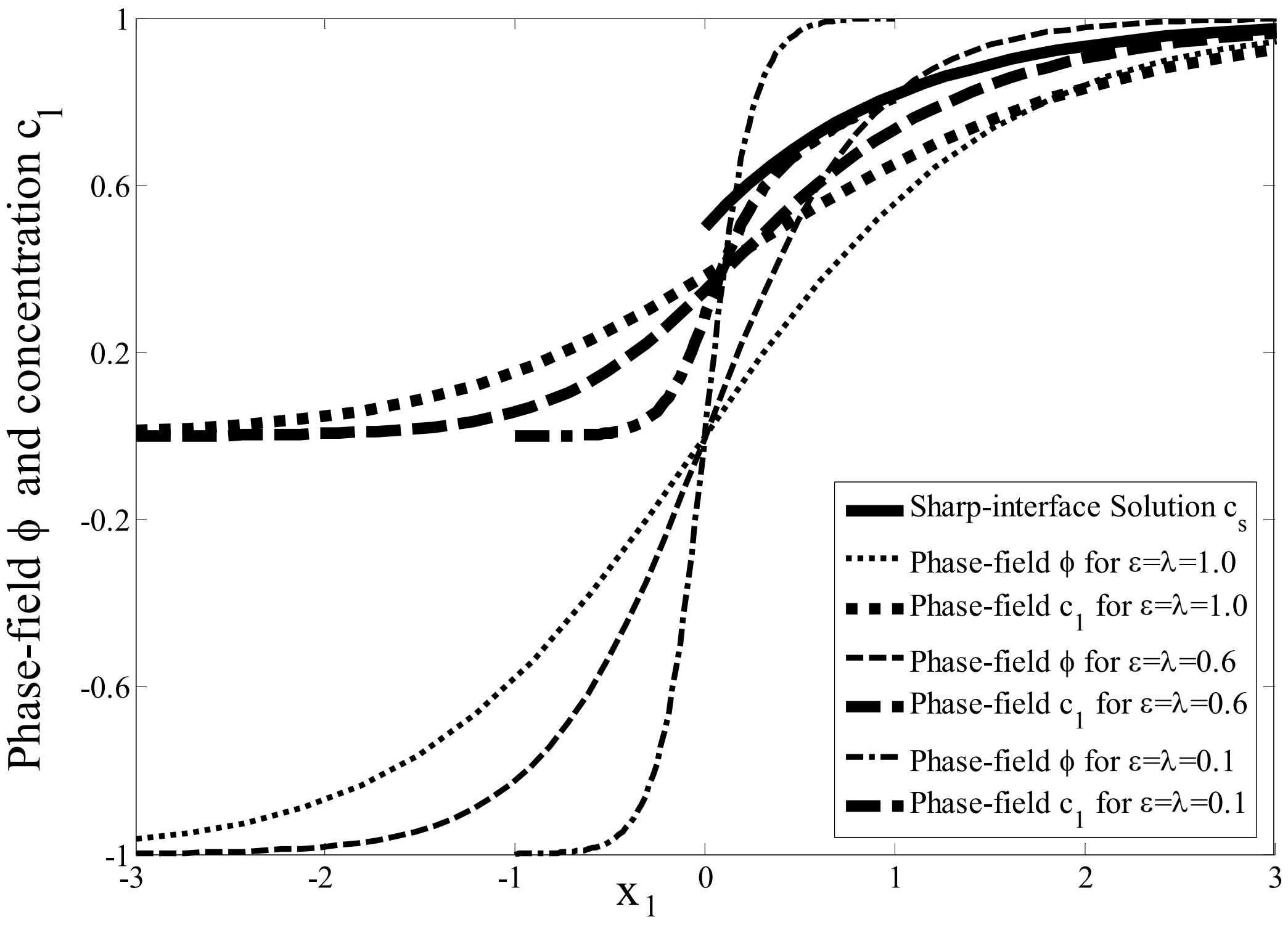

Phase-field ϕ and concentration $c_1$
$x_1$
1
0.6
0.2
-0.2
-0.6
-1
-3
-2
-1
0
1
2
3
Sharp-interface Solution $c_s$
Phase-field ϕ for ε=λ=1.0
Phase-field $c_1$ for ε=λ=1.0
Phase-field ϕ for ε=λ=0.6
Phase-field $c_1$ for ε=λ=0.6
Phase-field ϕ for ε=λ=0.1
Phase-field $c_1$ for ε=λ=0.1

FIG. 2.

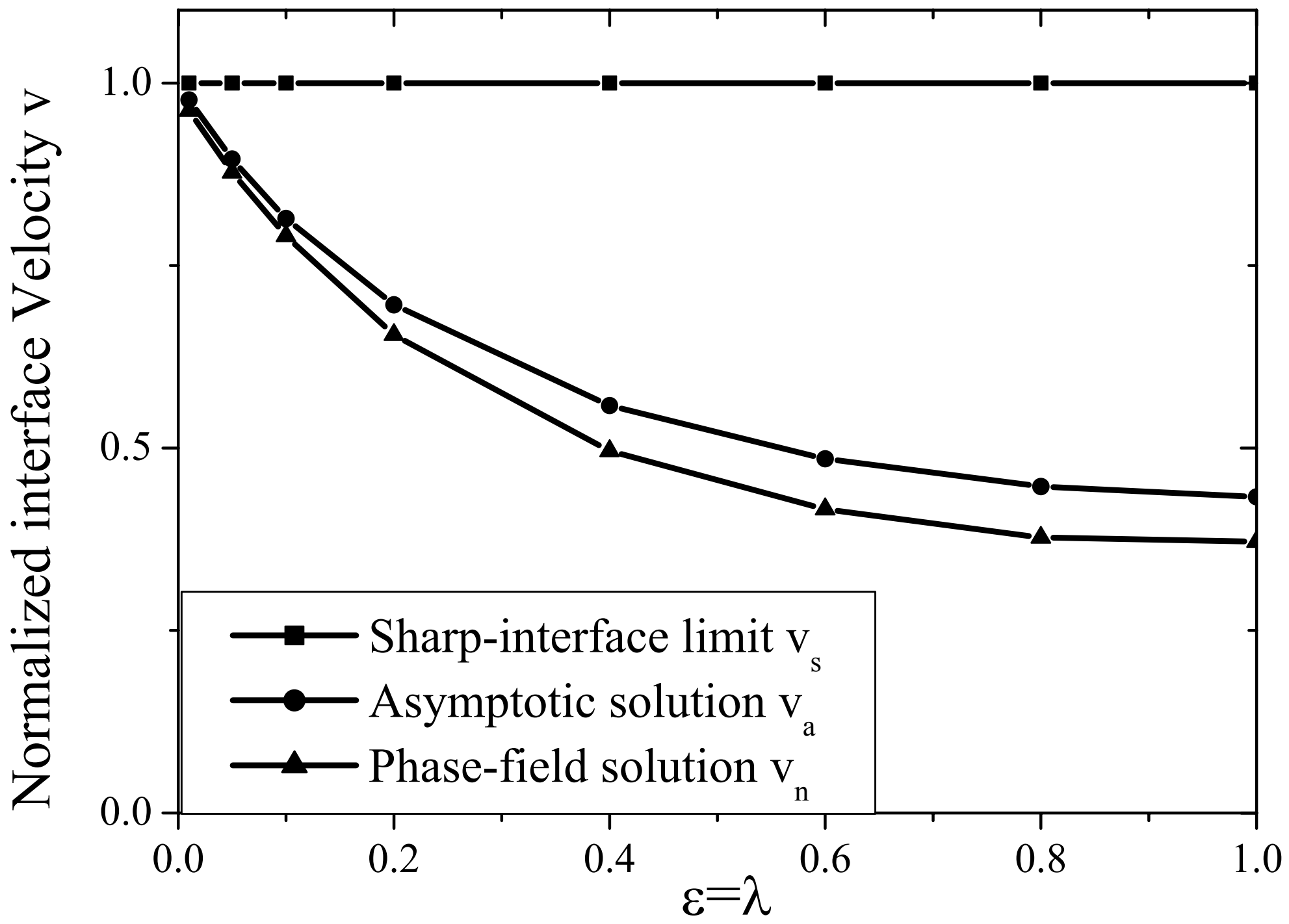

FIG. 3.